\documentclass[final]{cvpr}

\usepackage{times}
\usepackage{epsfig}
\usepackage{graphicx}
\usepackage{amsmath}
\usepackage{amssymb}
\usepackage{threeparttable}
\usepackage[section]{placeins}
\usepackage{multirow}
\usepackage{color}

\makeatletter
\def\hlinew#1{%
 \noalign{\ifnum0=`}\fi\hrule \@height #1 \futurelet
 \reserved@a\@xhline}
\makeatother%

\usepackage[pagebackref=true,breaklinks=true,colorlinks,bookmarks=false]{hyperref}

\def\cvprPaperID{} 
\def\confYear{}

\begin{document}

\title{Closing the Loop: Joint Rain Generation and Removal \\via Disentangled Image Translation}

\author{Yuntong Ye\textsuperscript{1,2}, Yi Chang\textsuperscript{2,}\footnotemark[1], Hanyu Zhou\textsuperscript{1}, Luxin Yan\textsuperscript{1}\\
\textsuperscript{1}National Key Laboratory of Science and Technology on Multispectral Information Processing,\\
School of Artificial Intelligence and Automation, Huazhong University of Science and Technology, China\\
\textsuperscript{2}AI Center, Peng Cheng Laboratory, Shenzhen, China\\
{\tt\small \{yuntongye, yichang, hyzhou, yanluxin\}@hust.edu.cn}
}

\maketitle

\pagestyle{empty}  
\thispagestyle{empty}

\begin{abstract}
	Existing deep learning-based image deraining methods have achieved promising performance for synthetic rainy images, typically rely on the pairs of sharp images and simulated rainy counterparts. However, these methods suffer from significant performance drop when facing the real rain, because of the huge gap between the simplified synthetic rain and the complex real rain. In this work, we argue that the rain generation and removal are the two sides of the same coin and should be tightly coupled. To close the loop, we propose to jointly learn real rain generation and removal procedure within a unified disentangled image translation framework. Specifically, we propose a bidirectional disentangled translation network, in which each unidirectional network contains two loops of joint rain generation and removal for both the real and synthetic rain image, respectively. Meanwhile, we enforce the disentanglement strategy by decomposing the rainy image into a clean background and rain layer (rain removal), in order to better preserve the identity background via both the cycle-consistency loss and adversarial loss, and ease the rain layer translating between the real and synthetic rainy image. A counterpart composition with the entanglement strategy is symmetrically applied for rain generation. Extensive experiments on synthetic and real-world rain datasets show the superiority of proposed method compared to state-of-the-arts.

\end{abstract}

~\\[-35pt]

\footnotetext[1]{Corresponding Author}
\section{Introduction}
\begin{figure}[t]
\begin{center}
   \includegraphics[width=1.0\linewidth]{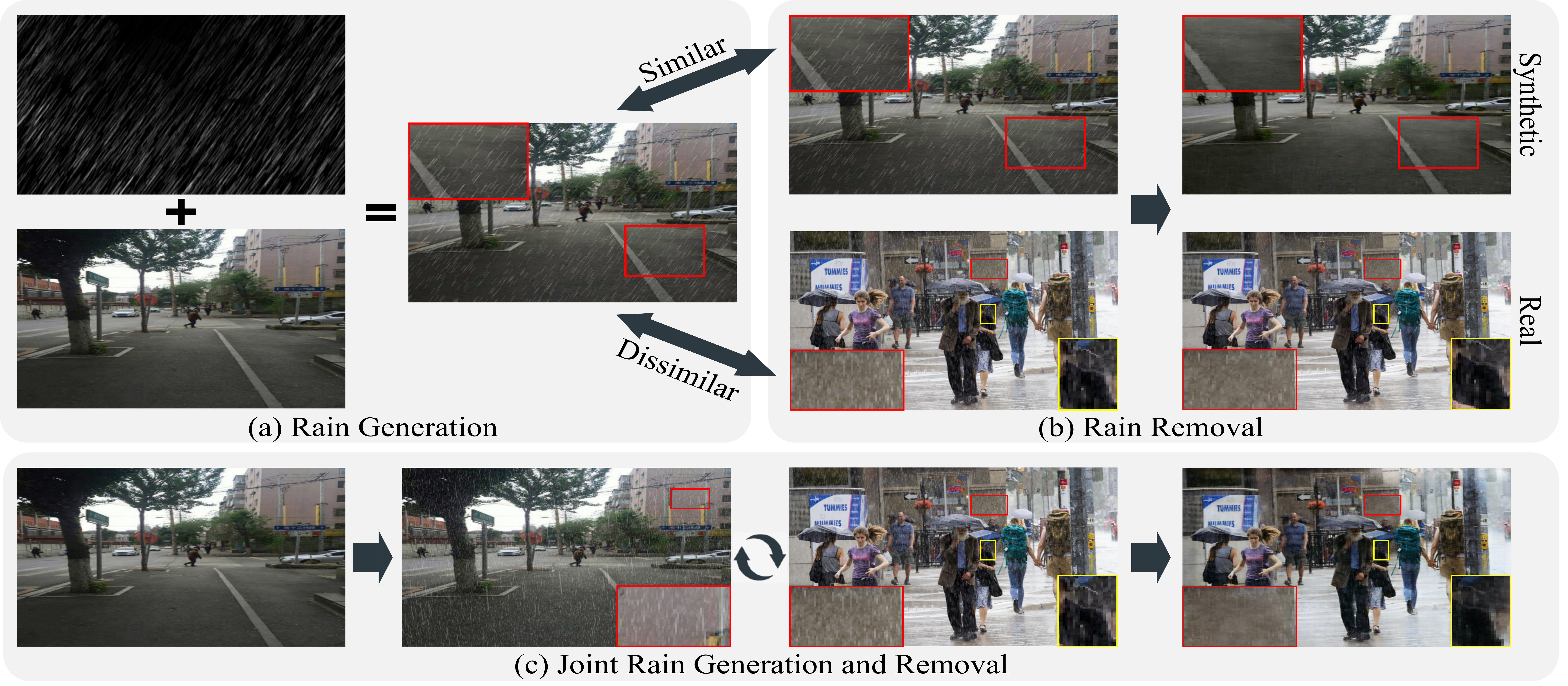}
\end{center}
      \caption{Illustration of rain generation and removal. (a) The hand-crafted simplified rain generation model. (b) Although the simulated rain has been well removed, the huge gap between synthetic training images and real-world testing images leads a significant performance drop in existing learning-based method JORDER \cite{yang2017deep}. (c) The proposed method learns the complex rain model from the data-driven perspective. We propose to jointly learn the rain generation and removal within a unified framework, as so to better bridges the domain gap between the real and synthetic rain.}
\label{Illustration}
\end{figure}

  Rain is a common weather phenomenon which dramatically degrades the quality of images and affects many computer vision tasks such as detection \cite{li2019single} and segmentation \cite{bahnsen2018rain}. The forward rain generation procedure \cite{li2016rain, yang2017deep, fu2017removing, hu2019depth, li2019heavy} is usually simplified as:
\begin{equation}
\setlength\abovedisplayskip{2pt}
\setlength\belowdisplayskip{2pt}
\textbf{\emph{O}} = \emph{\textbf{B}} + \textbf{\emph{R}},
\label{eq:addtion}
\end{equation}
  where $\textbf{\emph{O}}$, $\textbf{\emph{B}}$, $\textbf{\emph{R}}$ denote the rainy image, clean background and rain layer [Fig. \ref{Illustration}(a)]. Image deraining is formulated as an ill-posed inverse problem of the rain generation (1), aiming to recover the clean image \textbf{\emph{B}} from rainy image \textbf{\emph{O}}.

  Recently, the deep learning-based deraining methods have achieved remarkable performance, benefiting from powerful representation of convolutional neural network. Fu \emph{et al}. \cite{fu2017removing} introduced the end-to-end residual CNN for rain streaks removal task. Latter, more sophisticated CNN architectures \cite{yang2017deep, li2018recurrent, liu2018erase, wang2019erl, li2019heavy, yang2019scale, wang2020model, du2020conditional} have sprung up with tremendous progress, and the visual deraining results are impressive. Unfortunately, it is widely recognized that although the trained models can achieve satisfactory results on the synthetic rain, they cannot well generalize to the real rain [Fig. \ref{Illustration}(b)] because of the huge gap between the simplified synthetic rain and the complex real rain.

  To address this problem, a number of works have been proposed for better real rain removal. The first category starts from the intuitive rain generation perspective, in which the key idea of this research line is to make the simulated rain as real as possible in supervised manner. They tend to incorporate more complicated visual appearance of rain into consideration, such as  streaks, haze and occlusion degradation factors within a comprehensive rain model \cite{yang2017deep,  liu2018erase, li2019heavy, hu2019depth}. However, these hand-crafted generation models cannot well accommodate the complicated distribution of the real rain, due to the varied angle, location, depth, intensity, density, length, width and so on. Also, researchers try to generate the pair of the real rain and ‘clean’ image from the videos \cite{wang2019spatial} or the rendering technique \cite{halder2019physics}. On the one hand, the clean image generation is cumbersome; on the other hand, the generated clean image is still pseudo-label, not the oracle clean image. The second category directly resorts to the real image from the domain adaptation perspective \cite{wei2019semi, yasarla2020syn2real}. The key idea is to enforce additional constraint between the real and simulated deraining results. These semi-supervised/unsupervised methods can alleviate the domain gap by learning from the real rain, but they neglect the physical rain generation procedure.

  The previous methods either focus on the rain generation or  rain removal, few of them have noticed a simple yet importance problem that the rain generation and rain removal are of equal importance and should be tightly coupled. The rain removal is a typical inverse problem for the rain generation. A better rain generation model would definitely benefit for real rain removal, or vice versa shown in Fig. \ref{Illustration}(c).

  In this work, we bridge the gap between the rain generation and rain removal in an end-to-end learning framework. We bypass the difficulty of explicitly designing the sophisticated rain degradation model. Instead, our philosophy is to learn from real rainy image so as to approximate the real degradation implicitly. Specifically, we propose a bidirectional disentangled translation network [Fig. \ref{Framework}], in which each unidirectional network contains two stages of rain generation and removal for both the real and synthetic rain image, respectively. We observe that, in the image translation between the real rain image and simulated rain image, the background clean image layer is consistent while only the rain layer is changed. Instead of directly translating images from synthetic to real domain, this motivates us to preserve the identity in the image background while focus on transforming the simpler rain layer between the real ones and simulated ones. We employ both the self-consistency loss and adversarial loss for the image background. We summarize the main contributions as follows:

\begin{itemize}
\setlength{\itemsep}{-4pt}
\item We propose a novel image deraining algorithm which jointly learns the rain generation (forward) and rain removal (inverse) in a unified framework. Compared with the hand-crafted rain generation model, learning physical degradation from real rainy image could offer better approximation to the real rain in an implicit manner. Moreover, the rain generation and removal would benefit greatly from each other, thus improving the generalization for the real rainy images.
\item  We employ the disentanglement strategy in which the consistent background is well preserved by the self-consistency loss and adversarial loss. In contrast to previous methods which directly transform the simulated rain image to real rain image, the proposed method gets rid of the identity background and concentrates on the simpler rain layer translating, which significantly ease the difficulty of bridging the gap between real rain and simulated rain image.
\item We conduct extensive experiments on both synthetic and real-world datasets, which perform favorably against the state-of-the-art methods and consistently superior on real-world rainy images.
\end{itemize}

\section{Related Work}
\subsection{Rain Generation}
  Rain generation is the basis for the rain removal. Nayar and Garg \cite{garg2005does, garg2007vision} firstly developed a geometric and photometric model with detailed analysis about real rain appearance. Latter, the simple yet intuitive rain image decomposition model in  Eq. (\ref{eq:addtion}) has been widely used in the optimization-based methods \cite{li2016rain, zhu2017joint, chang2017transformed}. To better accommodate the complicated visual appearance of real rain, Yang \emph{et al}. \cite{yang2017deep} and Liu \emph{et al}. \cite{liu2018erase} extended the simple additive model (rain streaks) to the heavy rain model (distant veiling/haze effect) and occlusion rain model (close rain occlusions), respectively. Along this direction, Hu \emph{et al}. \cite{hu2019depth} have reached a very high level by taking both streaks, haze and occlusion degradation factors into a comprehensive rain model. Another research line tries to generate the rain image from the real rain. Wang \textit{et al.} \cite{wang2019spatial} proposed a semi-automatic method on rain sequence to generate paired high-quality clean-rainy images. Li \textit{et al.} \cite{li2019single} provided real rain dataset in driving and surveillance without clean counterparts. The rendering-based methods \cite{halder2019physics} have also been proposed to simulate the rain as real as possible.

  However, these complicated hand-crafted rain models still can not accurately reflect the raining procedure of the real physical world, due to the highly complex visual appearance of the real rain. In this work, compared with previous methods, we bypass the difficulty of explicitly designing the sophisticated rain degradation model. Instead, our philosophy is to learn from real rainy image so as to approximate the real degradation implicitly.

\begin{figure*}[h]

    \begin{minipage}[b]{1.0\linewidth}
      \centering
      \centerline{\includegraphics[width=17cm]{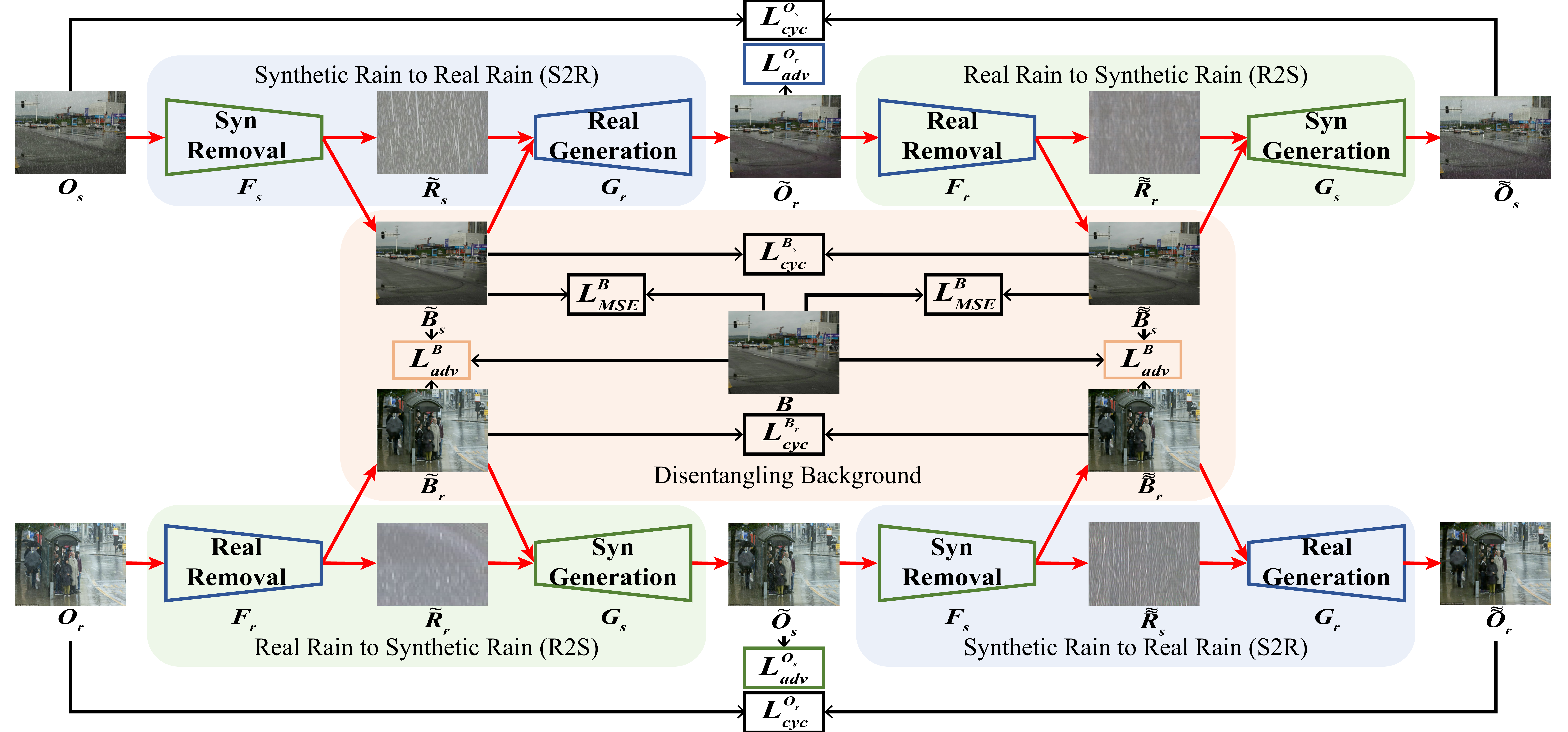}}
    \end{minipage}
        \caption{Overview of the proposed joint rain generation and removal (JRGR) framework. The proposed method consists of synthetic to real (S2R) and real to synthetic (R2S) modules, aiming to bridge the gap between synthetic and real rainy images. Instead of the directly translating, we embed the rain removal and generation into each module, which further improves the robustness of the proposed method. To better preserve the identity background image, we introduce both the self-consistency loss and adversarial loss for the clean background.}
    \label{Framework}
\end{figure*}

\subsection{Rain Removal}
  Rain removal is a highly ill-posed inverse problem to rain generation. Pioneer works design hand-crafted priors to decompose a rainy image into the background and rain layer \cite{kang2011automatic, chen2013generalized, luo2015removing, li2016rain, zhu2017joint}. Recently, the deep learning-based methods have achieved remarkable progress \cite{fu2017clearing, li2018recurrent, zhang2018density, wang2019dtdn, ren2019progressive, hu2019depth, li2019heavy, wei2019semi, yasarla2020syn2real, yasarla2019uncertainty}, benefiting from powerful representation of the CNN. Fu \emph{et al}. \cite{fu2017removing} firstly introduced an end-to-end residual CNN for rain streaks removal. Yang \textit{et al.} \cite{yang2017deep} jointly detected and removed the rain in a multi-task network. Hu \emph{et al}. \cite{hu2019depth} designed a depth-attention network to handle both the rain and haze. Albeit the impressive performance of the deraining methods on synthetic rain datasets, these methods suffer a significant performance drop in real-world scene due to the gap between synthetic training data and real testing data. To address this issue, utilizing both the synthetic paired data and real unpaired data, Wei \textit{et al.} \cite{wei2019semi} designed a semi-supervised framework via a weight-sharing network in which a Gaussian mixture model (GMM) is proposed to regularize the real rain layer. Inspired by \cite{wei2019semi}, Yasarla \textit{et al.} \cite{yasarla2020syn2real} further introduced Gaussian Process (GP) to model the real rain. In this work, we propose to jointly learn the rain generation and rain removal in disentangle image translation framework.

\subsection{Generative Adversarial Networks (GAN)}
  Recently, GAN has attracted extensive attention for image deraining due to its great power of synthesizing photo-realistic images \cite{qian2018attentive, zhang2019image, zhu2019singe, li2019heavy}. The generator learns the mapping from the rain image to the clean image, and then the discriminator tries to distinguish between fake/real samples during the train process. Recently, the CycleGAN \cite{zhu2017unpaired} based domain adaptation methods have been introduced for the real rain/haze removal with unpaired data \cite{DerainCycleGAN, shao2020domain}, aiming to translate the rainy/hazy image domain into the clean image domain in an unsupervised manner. The purposes are still on the rain/haze removal. Typically, Shao \textit{et al.} \cite{shao2020domain} utilized the architecture of CycleGAN + syn/real haze removal to generate hazy-clean pairs from unpaired real images, and finally focus on the dehazing. In this work, we emphasize the equal importance of rain removal and generation, in which the CycleGAN is formed as the internal cycles ($ \textup{removal} \leftrightarrow \textup{generation}$) within the external cycles ($\textup{synthetic} \leftrightarrow \textup{real rainy images}$) for the synergy of removal and generation. Moreover, we propose a novel disentanglement strategy by performing the translations on the rain layers instead of images, which significantly eases the translation since the rain space is much simpler than image space.

\section{Proposed Method}
\subsection{Framework Architecture}
  Given an real rainy image $\textbf{\emph{O}}$, our goal is to estimate the clean background $\textbf{\emph{B}}$. The huge gap between the synthetic training images $\textbf{\emph{O}}_\textbf{\emph{s}}$ and real testing image $\textbf{\emph{O}}_\textbf{\emph{r}}$ leads significant performance drop in full supervised learning-based methods. In this work, we propose to jointly learn real rain generation and removal procedure within a unified disentangled image translation framework, as shown in Fig. \ref{Framework}.

  Specifically, the overall architecture of the JRGR includes both external and internal cycles. The external cycles performs the synthetic and real cycle-consistency respectively, while the internal cycles performs the rain removal and generation cycle-consistency. On one hand, the proposed JRGR has constructed strong relationship between the real and synthetic rainy image via iteratively rain generation and removal (S2R and R2S modules). On the other hand, the overall architecture has great benefit for the feature propagation in both tasks. The S2R module translates the synthetic rainy image into a real rainy image via the disentanglement and entanglement, while R2S module performs the reverse process.

  Compared with previous direct image-to-image translation methods, the proposed method JRGR employs the disentangle strategy for better translation. The key importance of the disentanglement is how to choose a proper space with interpretability. The DRIT \cite{lee2018diverse} embeds the images onto two spaces: a domain-invariant content space capturing the shared information and a domain-specific attribute space. However, the two spaces are not clearly defined and constrained. On the contrary, JRGR disentangles the rainy image into the background and rain with clear physical meaning via both MSE and adversarial losses. Moreover, the conventional translation methods treat the two spaces equally, while we treat the spaces asymmetrically, in which the rain space is much simpler than image space. Not that, the rain layer is obtained by subtraction of rainy image and disentangled background. Thus we keep the background consistent while translate only between the real and synthetic rain space. Consequently, JRGR automatically exploits clean backgrounds, synthetic rain layers and real rain layers during the removal and generation, thus possessing the generalization ability in real-world deraining.

\subsection{Loss Functions}
In this section, we give a detail description about the loss functions. We apply unsupervised adversarial loss, self-supervised cycle-consistency loss and supervised MSE loss for each intermediate removal and generation output.

\noindent \textbf{Adversarial Loss.} To enforce the disentanglement of clean background which has no access to the ground truth, we employ a background discriminator ${{D}_\emph{B}}$ which attempts to distinguish the decomposed backgrounds and the photo-realistic background inputs provided by synthetic data. Meanwhile, the removal sub-network attempts to fool the discriminator by generating more realistic backgrounds. We impose the adversarial loss for removal sub-networks in both S2R and R2S modules:
\begin{equation}
\setlength{\abovedisplayskip}{3pt}
\setlength{\belowdisplayskip}{3pt}
\begin{aligned}
\mathcal{L}_{adv}^\emph{B}&({ {D}_{\emph{B}}, {F}_{\emph{s}}, {F}_{\emph{r}}})= \mathbb{E}_{{\emph{O}}_\textbf{\emph{s}}}[\textrm{log} (1- {{D}_{\emph{B}}}\left ({{F}_{\emph{s}}}(\textbf{\emph{O}}_\textbf{\emph{s}}) \right )) ]  +\\
& \mathbb{E}_{{\emph{O}}_{\emph{r}}}[ \textrm{log} (1-{ {D}_{\emph{B}}} \left ({{F}_{\emph{r}}}(\textbf{\emph{O}}_\textbf{\emph{r}}) \right )) ] + \alpha \mathbb{E}_{{\emph{B}}}[ \textrm{log} {{D}_\emph{B}}\left(\textbf{\emph{B}} \right)],
\label{eq2}
\end{aligned}
\end{equation}
where ${F_{\emph{s}}}$ and ${F_{\emph{r}}}$ denote synthetic and real rain removal sub-networks. $\alpha$ serves as the balance weight. In the cycles, another R2S and S2R modules are proposed on the generated synthetic and real rainy images $\tilde{\textbf{\emph{O}}}_\textbf{\emph{r}}$ and $\tilde{\textbf{\emph{O}}}_\textbf{\emph{s}}$ which form the closed-loop. Thus the background adversarial loss is further applied to the decomposed backgrounds $\tilde{\tilde{\textbf{\emph{B}}}}_\textbf{\emph{r}}$ and $\tilde{\tilde{\textbf{\emph{B}}}}_\textbf{\emph{s}}$. The final background adversarial loss is written as:
\begin{equation}
\setlength{\abovedisplayskip}{3pt}
\setlength{\belowdisplayskip}{3pt}
\begin{aligned}
&\resizebox{0.65\hsize}{!}{$\mathcal{L}_{adv}^B  {(D_{\emph{B}}, F_\textbf{\emph{s}}, F_{\emph{r}}, G_{\emph{s}}, G_{\emph{r}}) }= \alpha \mathbb{E}_{{\emph{B}}}[ \textrm{log} {D_{\emph{B}}} \left ( \textbf{\emph{B}} \right )]$}  \\
&\resizebox{0.905\hsize}{!}{$ + \mathbb{E}_{{\emph{O}}_{\emph{s}}}[\textrm{log} (1- {D_{\emph{B}}}\left ({F_{\emph{s}}}(\textbf{\emph{O}}_\textbf{\emph{s}}) \right )) + \textrm{log} (1 -{ D_{\emph{B}}}\left ({F_{\emph{r}}(G_{\emph{r}}(F_{\emph{s}}}(\textbf{\emph{O}}_\textbf{\emph{s}} ))) \right ))]$} \\
&\resizebox{0.905\hsize}{!}{$ + \mathbb{E}_{{\emph{O}}_{\emph{r}}}[\textrm{log} (1- {D_{\emph{B}}}\left ( {F_{\emph{r}}}(\textbf{\emph{O}}_\textbf{\emph{r}}) \right )) + \textrm{log} (1 -{D_{\emph{B}}}\left ( {F_{\emph{s}}(G_{\emph{s}}(F_{\emph{r}}}(\textbf{\emph{O}}_\textbf{\emph{r}} ))) \right ))]$}
\label{eq3}
\end{aligned}
\end{equation}
When the discriminator is updated, the background \textbf{\emph{B}} is randomly chosen in the former inputs, following the strategy in \cite{zhu2017unpaired}. By imposing Eq. \ref{eq3}, we constrain all the decomposed backgrounds in the same domain as the clean input $\textbf{\emph{B}}$.

After the disentanglement of backgrounds, we obtain the rain layers by subtraction. Since the gap between synthetic and real rainy images mainly exists in the rain layers, we perform the translation between rain layers by the rain generation sub-network. The translated rain layers are then entangled with the backgrounds to generate rainy images. For the generated synthetic and real images, we apply two rainy image discriminators to produce the adversarial losses. Taking the real rain generation as an example, the real rainy inputs $\textbf{\emph{O}}_\textbf{\emph{r}}$ are utilized to train the discriminator ${ D_{O_r}}$ and generation sub-network ${ G_r}$ in an adversarial manner:
\begin{equation}
\setlength{\abovedisplayskip}{3pt}
\setlength{\belowdisplayskip}{3pt}
\begin{aligned}
\mathcal{L}_{adv}^{O_r} & {(D_{O_{\emph{r}}}, F_{\emph{s}}, G_{\emph{r}})} = \mathbb{E}_{{\emph{O}}_{\emph{r}}}[ \textrm{log} {D_{O_{\emph{r}}}}\left ( \textbf{\emph{O}}_\textbf{\emph{r}} \right )] \\
&+ \mathbb{E}_{{\emph{O}}_{\emph{s}}}[ \textrm{log} {(1 - D_{O_\textbf{\emph{r}}}}\left ( {G_{\emph{r}}(F_{\emph{s}}}(\textbf{\emph{O}}_\textbf{\emph{s}})) \right ))],
\end{aligned}
\end{equation}
A similar adversarial loss $\mathit{L}_{adv}^{O_s}$ is imposed on synthetic generation sub-network ${G_{\emph{s}}}$ and the discriminator ${D_{O_{\emph{s}}}}$ to enforce the generation of synthetic rainy image $\tilde{\textbf{\emph{O}}}_\textbf{\emph{s}}$.

\noindent \textbf{Cycle-consistency Loss.} Although the adversarial losses guarantee the generated images and the target images in the same domain, the content information in rainy images may be degraded during the generation. Benefiting from the closed-loop in our framework, we employ four cycle-consistency losses which self-supervise the removal and generation sub-networks to preserve the content information. Firstly, while S2R module generates real rainy images from synthetic rainy images, the reconstructed synthetic output which is generated by the latter reverse process of R2S module should be consistent with the synthetic input. Thus we define the cycle-consistency loss as:
\begin{equation}
\setlength{\abovedisplayskip}{3pt}
\setlength{\belowdisplayskip}{3pt}
\begin{aligned}
\resizebox{0.9\hsize}{!}{$\mathcal{L}_{cyc}^{O_s} {(F_{\emph{s}}, F_{\emph{r}}, G_{\emph{s}}, G_{\emph{r}})} = \mathbb{E}_{{\emph{O}}_{\emph{s}}}[\|\textbf{\emph{O}}_\textbf{\emph{s}}-{G_{\emph{s}}(F_{\emph{r}}(G_{\emph{r}}(F_{\emph{s}}}(\textbf{\emph{O}}_\textbf{\emph{s}})))) \|_1]$.}
\end{aligned}
\end{equation}
Meanwhile, for the reconstruction of $\textbf{\emph{O}}_\textbf{\emph{r}}$, a similar cycle-consistency loss $\mathit{L}_{cyc}^{O_r}$ is applied to train the other cycle.

Secondly, the generated real rainy images $\tilde{\textbf{\emph{O}}}_\textbf{\emph{r}}$ and the original synthetic rainy images $\textbf{\emph{O}}_\textbf{\emph{s}}$ should possess consistent backgrounds. The cycle-consistency loss is written as:
\begin{equation}
\setlength{\abovedisplayskip}{3pt}
\setlength{\belowdisplayskip}{3pt}
\begin{aligned}
\resizebox{0.9\hsize}{!}{$\mathcal{L}_{cyc}^{B_s}{(F_{\emph{s}}, F_{\emph{r}}, G_{\emph{r}})} = \mathbb{E}_{{\emph{O}}_{\emph{s}}}[\| {F_{\emph{s}}}(\textbf{\emph{O}}_\textbf{\emph{s}})-{(F_{\emph{r}}(G_{\emph{r}}(F_{\emph{s}}}(\textbf{\emph{O}}_\textbf{\emph{s}}))))\|_1]$.}
\end{aligned}
\end{equation}
In a similar fashion, the cycle-consistency loss $\mathit{L}_{cyc}^{B_r}$ is defined for the decomposed backgrounds $\tilde{\textbf{\emph{B}}}_\textbf{\emph{r}}$ and $\tilde{\tilde{\textbf{\emph{B}}}}_\textbf{\emph{r}}$.

\noindent \textbf{MSE Loss.} In addition to the adversarial losses and the cycle-consistency losses, for the synthetic rainy images which have ground truth background counterparts, we utilize the MSE loss to supervise the training of synthetic removal sub-network, which is written as:
\begin{equation}
\setlength{\abovedisplayskip}{4pt}
\setlength{\belowdisplayskip}{4pt}
\begin{aligned}
\mathcal{L}_{mse}^{B_s} {(F_{\emph{s}}, F_{\emph{r}}, G_{\emph{r}})} &= \| {F_{\emph{s}}}(\textbf{\emph{O}}_\textbf{\emph{s}})- \textbf{\emph{B}} \|^2_2 \\
&+ \| {F_{\emph{r}}(G_{\emph{r}}(F_{\emph{s}}}(\textbf{\emph{O}}_\textbf{\emph{s}}))) - \textbf{\emph{B}} \|^2_2.
\end{aligned}
\end{equation}

\noindent \textbf{Full Objective.} The full objective function contains the adversarial loss, cycle-consistency loss, and the MSE loss as follow:
\begin{equation}
\begin{aligned}
\setlength{\abovedisplayskip}{3pt}
\setlength{\belowdisplayskip}{3pt}
&\resizebox{0.95\hsize}{!}{$\mathcal{L}({ F_{\emph{s}}, F_{\emph{r}}, G_{\emph{s}}, G_{\emph{r}}, D_{\emph{B}},{D_{O_{\emph{s}}}},{D_{O_{\emph{r}}}}})  = \lambda_{adv}(\mathit{L}_{adv}^{B} + \mathit{L}_{adv}^{D_{O_s}} + \mathit{L}_{adv}^{D_{O_r}})$} \\
& +\lambda_{cyc}(\mathit{L}_{cyc}^{O_r} + \mathit{L}_{cyc}^{O_s} + \mathit{L}_{cyc}^{B_r} + \mathit{L}_{cyc}^{B_s}) +\lambda_{mse}\mathit{L}_{mse}^{B_s}.
\end{aligned}
\end{equation}
By imposing the full objective function, we alleviate the problem that paired real rainy-clean images are not accessible. In our framework, each removal and generation sub-network is constrained by at least one unsupervised adversarial loss and one self-supervised cycle-consistency loss to guarantee the domain of the outputs while preserve the content information, which consequently guides the framework to exploit the clean backgrounds and rain layers.

\subsection{Implementation Details}
The framework is implemented using the PyTorch with four RTX 2080Ti GPUs. We utilize the U-Net \cite{ronneberger2015u} as the removal and rain generation sub-networks. Subtraction and summation are followed to obtain the disentangled rain layers and generated rainy images. The PatchGAN \cite{isola2017image, ledig2017photo} is utilized to construct the discriminators. We empirically set the balance weight $\alpha$, $\lambda_{adv}$, $\lambda_{cyc}$ and $\lambda_{mse}$ as 4, 10, 1, 10.

The synthetic and real images are randomly cropped into $256 \times 256$ as input of each sub-network. We first pre-train the synthetic removal sub-network and real removal sub-network on paired synthetic dataset for 100 epochs with the learning rate 0.0001. Then we jointly train JRGR using paired synthetic data and unpaired real data for 200 epochs, in which the learning rates of generation sub-networks are set as 0.0001 while divided by 10 and 100 for real and synthetic removal sub-network. The Adam optimizer is adopted with batch size 16. In testing, we apply the real removal sub-network to output clean backgrounds.

\section{Experiments}

\subsection{Datasets and Experimental Setting}
\noindent \textbf{Datasets.} We conduct experiments on both synthetic and real rain datasets to evaluate the proposed method.
~\\[-18pt]
\begin{itemize}
\setlength{\itemsep}{-4pt}
\item \textbf{Cityscape.} We utilize two different synthetic rain datasets, RainCityscape \cite{hu2019depth} and Rendering \cite{halder2019physics}, which synthesize rain and haze with different models on the Cityscape dataset \cite{cordts2016cityscapes}. To simulate the real situation, during training, we use 1400 paired rainy-clean images in RainCityscape as synthetic data and 1400 unpaired rainy images in RainRendering with no access to the clean counterparts as real data. 175 rainy images in RainRendering are used for testing.

\begin{figure*}[h]
    \begin{minipage}[b]{1.0\linewidth}
      \centering
      \centerline{\includegraphics[width=17cm]{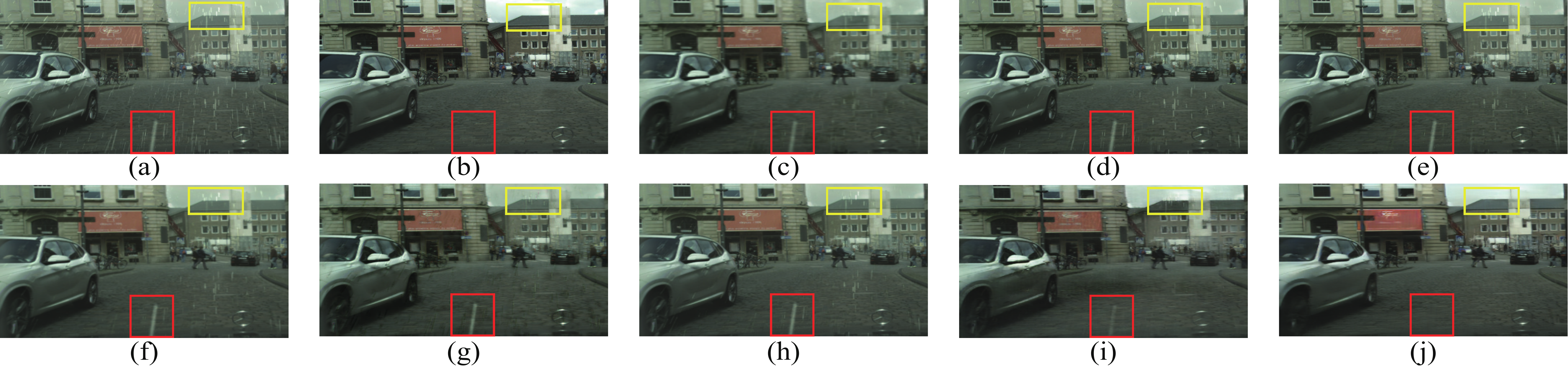}}
    \end{minipage}
    \caption{Visualization of deraining results on Rendering dataset. (a) Rainy image. (b) Clean image. Deraining results by (c) DDN, (d) JORDER-E, (e) RESCAN (f) DAF, (g) SSIR, (h) Syn2Real, (i) Cycle GAN, (j) JRGR.}
    \label{RenderingResult}
\end{figure*}

\item \textbf{RainHQ.} RainHQ dataset is a high quality real rain dataset which provides paired rainy-clean patches via semi-automatic labelling methods \cite{wang2019spatial}. We use 2000 paired synthetic rainy-clean images following the strategy in \cite{fu2017removing} and 2000 unpaired rainy images for training. 500 real rainy images are utilized for testing.

\item \textbf{RealRain.} We also collect unpaired real world rainy images and clean backgrounds with large field of view from the datasets provided by \cite{yang2017deep, wei2017should, zhang2019image, chen2018robust} and Google search. We utilize totally 400 real rainy images in RealRain as the unpaired rainy images and 400 paired synthetic rainy images to train the framework. 88 real rainy images are utilized for test.
\end{itemize}

\noindent \textbf{Experimental Setting.} We compare JRGR with (1) supervised deraining methods DDN \cite{fu2017removing}, JORDER-E \cite{yang2019joint}, RESCAN \cite{li2018recurrent}, DAF \cite{hu2019depth} and the GAN-based generation method pix2pix \cite{isola2017image}, which are trained with paired synthetic rainy-clean images and tested in real rainy images; (2) unsupervised method Cycle GAN \cite{zhu2017unpaired} trained with real rainy images and clean backgrounds provided by synthetic data; (3) semi-supervised methods SIRR \cite{wei2019semi} and Syn2Real \cite{yasarla2020syn2real} which are trained with both synthetic and real data. PSNR and SSIM are utilized for quantitative evaluation.

\begin{figure*}[t]
    \begin{minipage}[b]{1.0\linewidth}
      \centering
      \centerline{\includegraphics[width=17cm]{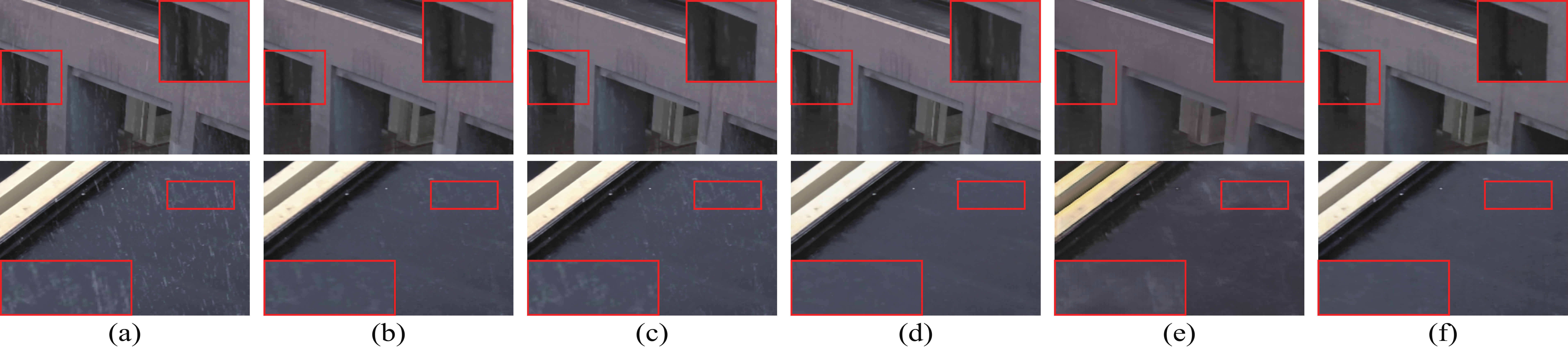}}
    \end{minipage}
    \caption{Visualization of deraining results on RainHQ dataset. (a) Rainy image. Deraining results by (b) DDN, (c) JORDER-E, (d) Syn2Real, (e) Cycle GAN, (f) JRGR.}
    \label{RainHQResult}
\end{figure*}

\begin{figure*}[t]
    \begin{minipage}[b]{1.0\linewidth}
      \centering
      \centerline{\includegraphics[width=17cm]{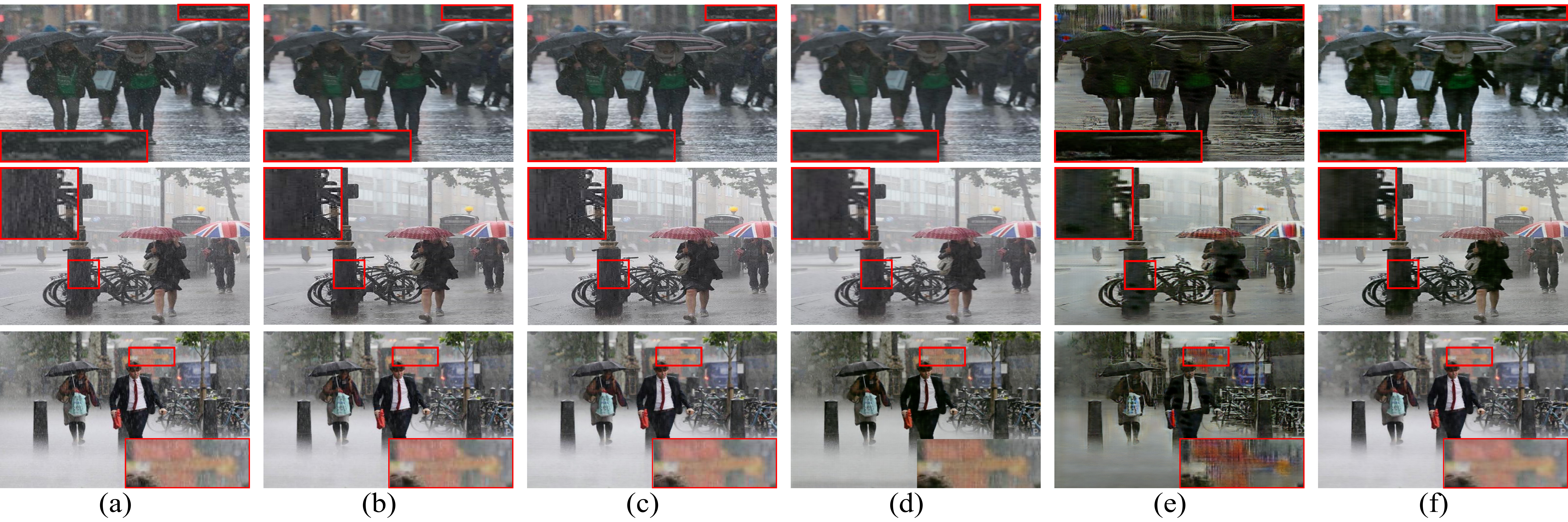}}
    \end{minipage}
    \caption{Visualization of deraining results on RealRain dataset. (a) Rainy image. Deraining results by (b) DDN, (c) JORDER-E, (d) Syn2Real, (e) Cycle GAN, (f) JRGR.}
    \label{RealResult}
\end{figure*}

\begin{figure*}[t]
    \begin{minipage}[b]{1.0\linewidth}
      \centering
      \centerline{\includegraphics[width=1.0\linewidth]{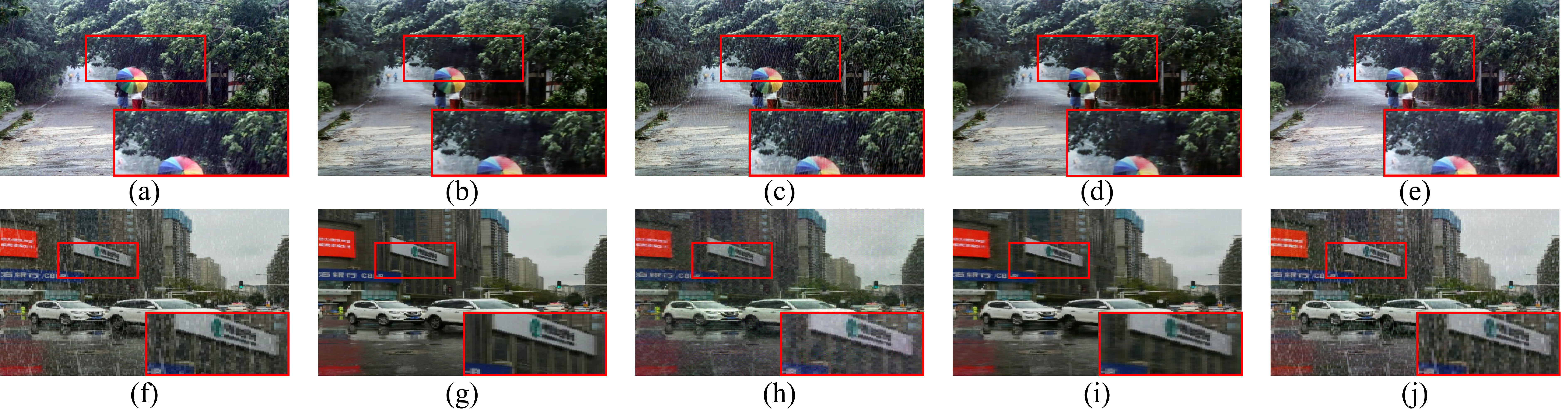}}
    \end{minipage}
    \caption{The effectiveness of the joint removal-generation framework. (a) Real input, (b) decomposed background from (a), (c) generated synthetic rainy image, (d) decomposed background from (c), (e) reconstructed real input, (f) synthetic input, (g) decomposed background from (f), (h) generated real rainy image, (i) decomposed background from (h), (g) reconstructed synthetic input. By the mutual effect of removal and generation sub-networks, JRGR obtains good performance in each intermediate output.}
    \label{Components}
\end{figure*}

\subsection{Experiments on Synthetic Images}

Figure \ref{RenderingResult} shows the visual results on synthetic dataset Rendering \cite{halder2019physics}. For the supervised methods, they cannot well generalize on Rendering dataset due to the different rain streak appearances in training set RainCityscape and testing set Rendering. As a result, the rain streaks exist in most deraining results. SSIR learns the rain streaks in Rendering in a semi-supervised manner and removes the rain streaks in the sky. However, the hand-craft designed GMM is limited to model the diverse rain such as the long rain streak on the floor. The unsupervised Cycle GAN generates artifacts on the floor which degrades the visualization quality in deraining result. The proposed JRGR learns the distribution of backgrounds and rain layers during removal and generation, which consequently generates the satisfactory deraining results. The quantitative results in Table \ref{MethodCompare} also demonstrate the superiority of our proposed method.

\begin{table}[t]
  \centering
  \caption{Quantitative comparison with state-of-art methods on synthetic and real datasets.}
  \begin{threeparttable}
  \setlength{\tabcolsep}{2.0mm}{
  \begin{tabular}{c|cc|cc}
   \hlinew{2pt}
   \multirow{2}{*}{Dataset} & \multicolumn{2}{c|}{Rendering \cite{halder2019physics}} & \multicolumn{2}{c}{RainHQ \cite{wang2019spatial}} \\
   \cline{2-5}
    & PSNR & SSIM & PSNR & SSIM \\
   \hlinew{2pt}
    DDN \cite{fu2017removing} & 24.12 &  0.8781 & 33.74 & 0.9112\\
    RESCAN \cite{li2018recurrent} & 24.89& 0.9101 & 33.11 & 0.9253\\
    DAF \cite{hu2019depth} & 25.23& 0.8827 & 24.15 & 0.8523 \\
    JORDER-E \cite{yang2019joint} & 25.64& 0.8767 & 33.28 & 0.9406\\
    pix2pix \cite{isola2017image} & 25.21& 0.8833 & 32.43 & 0.9148 \\
    Cycle GAN \cite{zhu2017unpaired} & 26.29& 0.8921 & 33.54 & 0.9127  \\
    SSIR \cite{wei2019semi} & 25.08& 0.8853 &30.78 & 0.8668 \\
    Syn2Real \cite{yasarla2020syn2real} & 25.32& 0.8871 & 33.14 & 0.9183\\
    JRGR & \textbf{27.51}& \textbf{0.9132} & \textbf{35.59} & \textbf{0.9498}\\
   \hlinew{2pt}

  \end{tabular}}
  \end{threeparttable}
  \label{MethodCompare}
 \end{table}

\subsection{Experiments on Real Images}
In Fig. \ref{RainHQResult}, we visualize the rain removal results on real rain dataset RainHQ \cite{wang2019spatial}, in which the rain is more complex than Rendering. State-of-the-art deraining methods have difficulty in dealing with real-world rain and leave the rain streaks unremoved. The artifacts in Cycle GAN degrade deraining results. The proposed method possesses the clean and smooth backgrounds. Quantitative comparison is shown in Table \ref{MethodCompare}. The performance of DAF \cite{hu2019depth} significantly drops without re-training using depth. Cycle GAN possesses a comparable quantitative result. By imposing disentanglement and translation on the rain layer, the proposed method obtains more natural results and achieves the best performance in terms of PSNR and SSIM.

\begin{table}[htb]
  \centering
  \caption{The ablation study for different loss functions and training strategies on real rain dataset RainHQ. Init-1 and Init-2 denote two different training strategies.}
  \begin{threeparttable}
  \setlength{\tabcolsep}{8mm}{
  \begin{tabular}{c|cc}
   \hlinew{2pt}
    Strategy & PSNR & SSIM\\
   \hlinew{2pt}
	 w/o $\mathit{L}_{adv}^B$  & 27.38 & 0.8926\\
	 w/o $\mathit{L}_{adv}^O$ & 28.15& 0.9012\\
	 w/o $\mathit{L}_{cyc}$  & 26.87 & 0.8902\\
	 w/o $\mathit{L}_{MSE}$  & 31.24 & 0.9144\\
	 Init-1 & 33.97 & 0.9305 \\
	 Init-2 & 34.54 & 0.9424 \\
	 Proposed & \textbf{35.59} & \textbf{0.9498} \\
   \hlinew{2pt}
  \end{tabular}}
  \end{threeparttable}
  \label{ablationTab}
\end{table}

Furthermore, we show the visual results on RealRain dataset in Fig. \ref{RealResult}. DDN, JORDER-E and Syn2Real tend to leave the rain streak on deraining results. More artifacts are produced by Cycle GAN when facing the complex scene. The proposed JRGR removes most of the rain, while preserves the details of backgrounds.

\subsection{Ablation Study}
In this section, ablation study on RainHQ is conducted to evaluate the effectiveness of losses and training strategy.

\noindent \textbf{Effectiveness of Losses.} Denoting $\mathit{L}_{adv}^O$ as the rainy image adversarial losses, i.e., $\mathit{L}_{adv}^{O_s}$ and $\mathit{L}_{adv}^{O_t}$, while $\mathit{L}_{cyc}$ as the cycle-consistency losses, we investigate the effectiveness of the loss functions in Table \ref{ablationTab}. The performance drops in the first four rows demonstrate the benefits brought by the adversarial, cycle-consistency and MSE loss.

\noindent \textbf{Effectiveness of Training Strategy.} We also study the effectiveness of training strategy in Table \ref{ablationTab}. Init-1 trains the sub-networks in JRGR together with no pre-training. Init-2 pre-trains the synthetic removal sub-network on the paired synthetic dataset and then trains the sub-networks together, in which the learning rate of synthetic removal sub-network is divided by 100. The proposed method further pre-trains the real removal sub-network and then trains the sub-networks together, in which the learning rate of real removal sub-network is divided by 10. The quantitative results show the initialization of the synthetic and real removal sub-networks is necessary.

\subsection{Discussion}

\begin{figure}[t]
    \begin{minipage}[b]{1.0\linewidth}
      \centering
      \centerline{\includegraphics[width=1.0\linewidth]{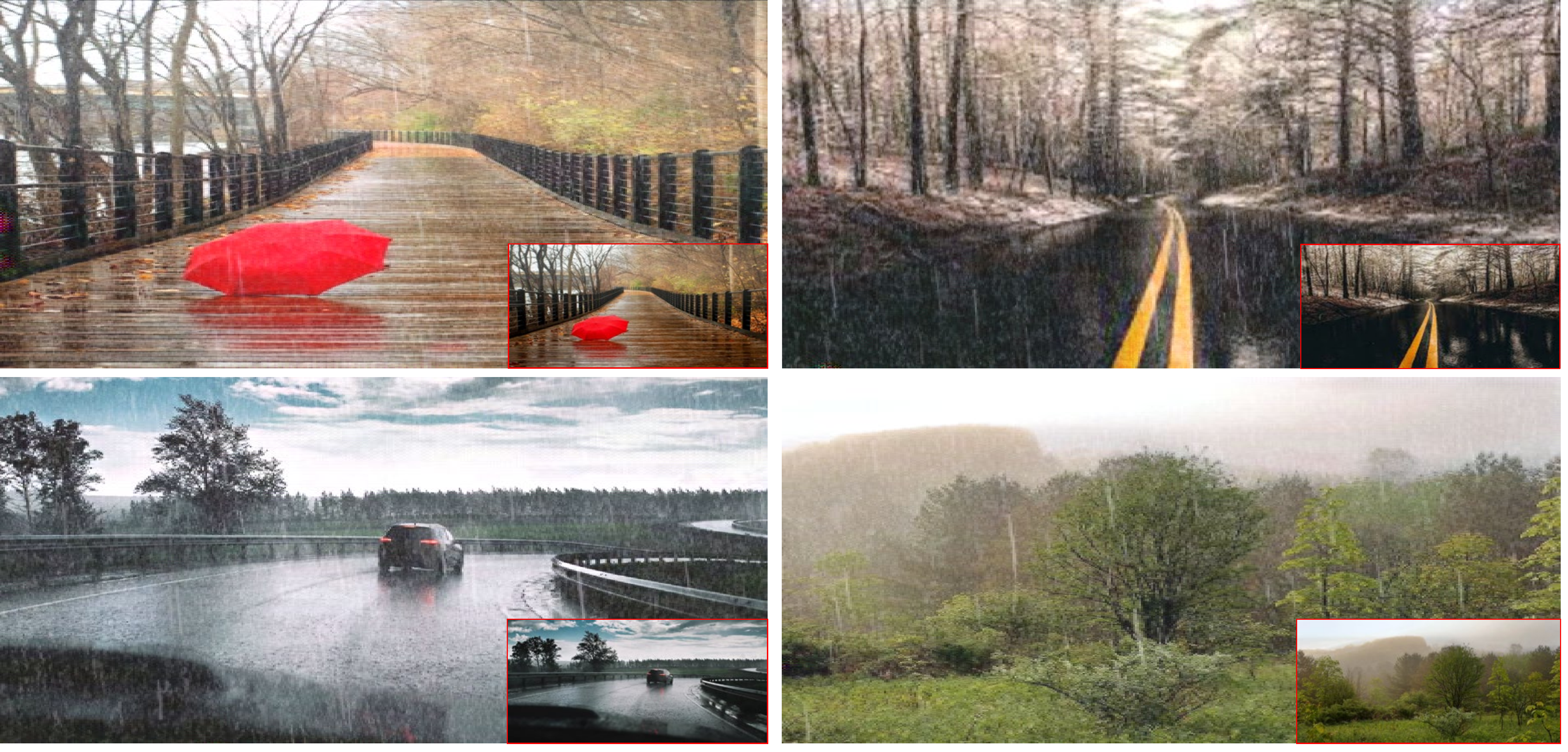}}
    \end{minipage}
    \caption{Visualization results of generated real rains by JRGR.}
    \label{QuanGen}
\end{figure}

\noindent  \textbf{Analysis of Joint Rain Removal and Generation.} We investigate all the intermediate removal and generation output of JRGR. In Fig. \ref{Components}, all the estimated backgrounds (b), (d), (g), (i) are clean without rain streaks. The rain streaks in generated synthetic rainy image are in a synthetic style [Fig. \ref{Components}(c)], while the generated real rainy image [Fig. \ref{Components}(h)] possesses more photo-realistic rain which is similar with the input [Fig. \ref{Components}(a)]. The final outputs [(Fig. \ref{Components}(e),  (j)] successfully reconstruct the inputs [Fig. \ref{Components}(a), (f)]. The proposed JRGR could provide physical meaning results with good performance via the adversarial and cycle-consistency losses which guarantee both the translation performance and identity preservation. In Fig. \ref{QuanGen}, we present more high quality realistic rainy images generated by JRGR, which includes both the natural rain veiling and streaks.

\begin{figure}[t]
    \begin{minipage}[b]{1.0\linewidth}
      \centering
      \centerline{\includegraphics[width=1.0\linewidth]{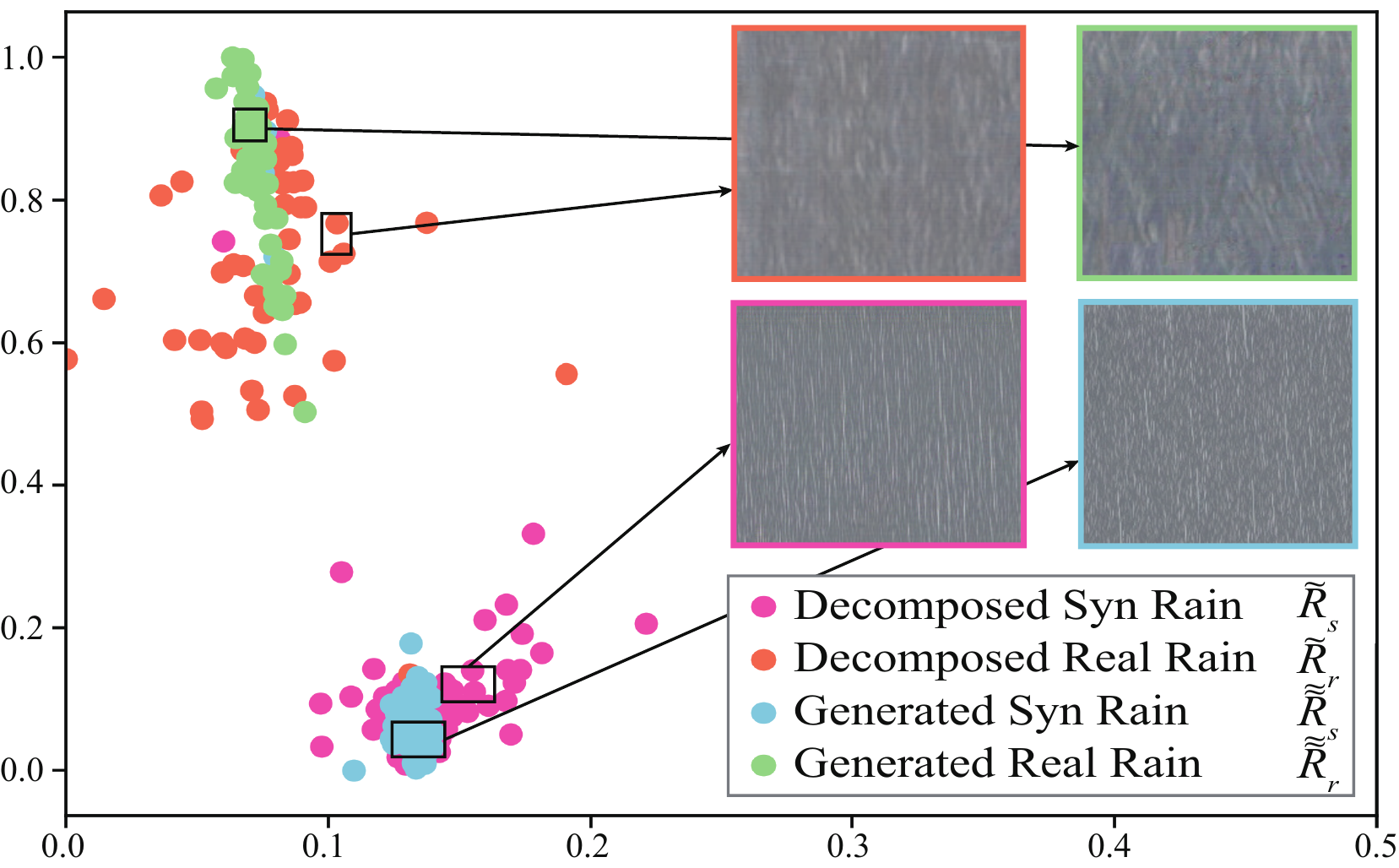}}
    \end{minipage}
    \caption{Visualization results of t-SNE on the decomposed and generated rain layers in real and synthetic data.}
    \label{tSNE}
\end{figure}

\noindent \textbf{Analysis of Rain Layer Translation.} To study the performance of rain layer translation, we visualize the features of synthetic and real rain layers using t-SNE in Fig. \ref{tSNE}. On the one hand, the separation of the synthetic and real rain layer feature demonstrates the huge gap in synthetic and real rain, which is also illustrated by the appearance difference of the corresponding rain layer visualizations. On the other hand, the decomposed and generated rain layers are aligned in both synthetic and real rain domain (green with orange, blue with pink). The visualized decomposed and generated rain layers are also similar, which demonstrates that our rain generators successfully perform the translation between synthetic and real rain layers.

\begin{figure}[t]
    \begin{minipage}[b]{1.0\linewidth}
      \centering
      \centerline{\includegraphics[width=1.0\linewidth]{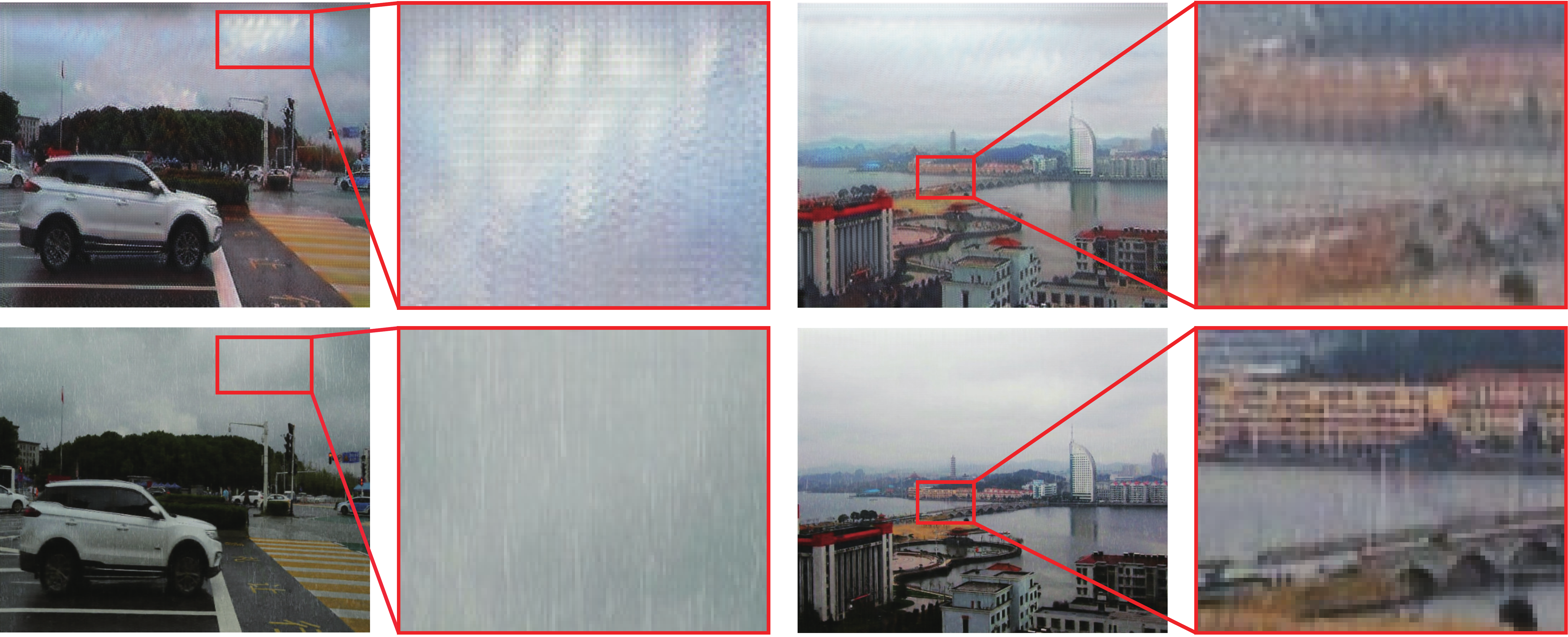}}
    \end{minipage}
    \caption{The superiority of translation between rain layers. Top: CycleGAN directly translates background into rainy image leaving obvious artifact. Bottom: JRGR preserves the background and focuses on the translation of rain layer, obtaining natural result.}
    \label{CycleCompare}
\end{figure}

Furthermore, to show the superiority of performing translation between simpler rain layers, we compare JRGR with Cycle GAN which directly performs the translation between backgrounds and rainy images in Fig. \ref{CycleCompare}. Cycle GAN pays attention to the most discriminative features between backgrounds and rainy images, which may not be the rain but the other factors such as textures and colors. On the contrary, the backgrounds of JRGR are well decomposed and preserved during removal, which makes it possible for the generator to focus on the translation between rain layers. That is the main reason why the JRGR could obtain more natural results than that of the Cycle GAN.

\begin{figure}[t]
    \begin{minipage}[b]{1.0\linewidth}
      \centering
      \centerline{\includegraphics[width=1.0\linewidth]{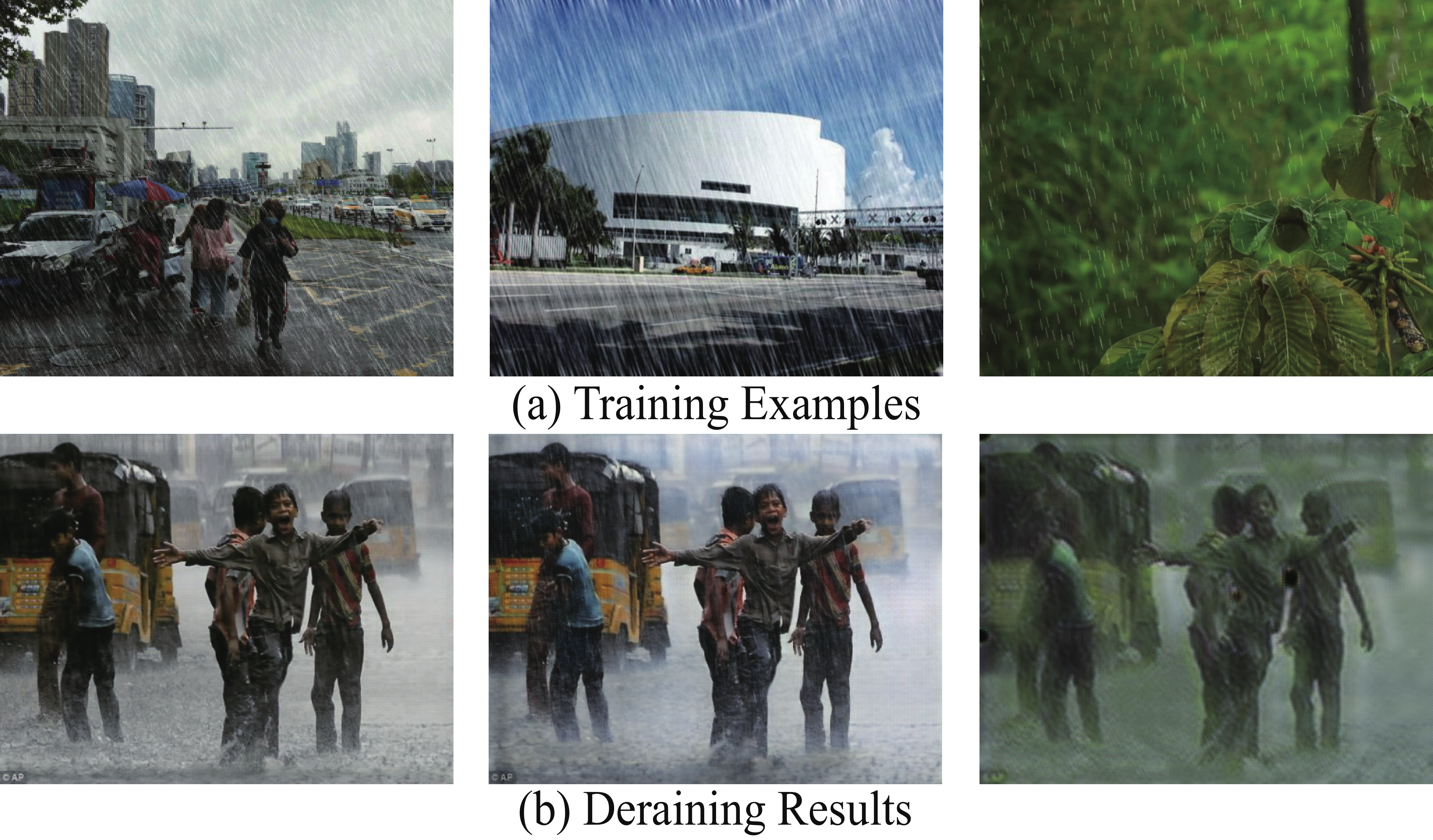}}
    \end{minipage}
    \caption{The limitation of our JRGR for intra-domain gap. (a) Examples of backgrounds in three different training synthetic datasets. (b) JRGR tends to generate results with different color styles when trained with different datasets. JRGR fails when the training and testing backgrounds are significantly different.}
    \label{limitation}

\end{figure}

\noindent \textbf{Limitation.} The JRGR mainly bridges the inter-domain gap between synthetic and real rainy images. However, there still exists the
intra-domain gap within real rainy images. For example, the real rainy city and forest images contain the different color and texture characteristics. These intra-domain gaps in the real rainy image would lead to the performance drop. In Fig. \ref{limitation}, we show the deraining examples by three datasets with different backgrounds, i.e., city under overcast sky, city under clear sky and forests. When the backgrounds behind real and synthetic data are significantly different, the proposed method fails to generate good deraining results. For example, the method trained by overcast sky dataset obtains the best performance for the backgrounds are most similar with the rainy images. Definitely, the comprehensive datasets would alleviate the problem from the dataset perspective. We would like to tackle the inter and intra-domain gaps simultaneously in future.

\section{Conclusion}
In this paper, we have proposed a disentangled image translation framework which jointly learns rain generation and removal to address the real-world deraining problem. Specifically, a bidirectional translation network are constructed by the removal and generation sub-networks with tightly coupled generation and removal modules. We further employ the disentangled strategy on the rain image by decomposing the rainy image into clean background and rain layer, so as to preserve the identity background and ease the translation with only the rain layer. The removal and generation sub-networks, constrained by the adversarial and cycle-consistency losses, mutually affect each other and consequently endow the framework with the generalization ability of dealing with real rain. The extensive experiments on synthetic and real rain datasets demonstrate the superiority of the proposed framework.

\textbf{Acknowledgements.} This work was supported by National Natural Science Foundation of China under Grant No. 61971460, China Postdoctoral Science Foundation under Grant 2020M672748, National Postdoctoral Program for Innovative Talents BX20200173 and Industrial Technology Development Program grant JCKY2018204B068.

{\small
\bibliographystyle{ieee_fullname}
\bibliography{egbib}
}

\end{document}